\documentstyle[aps,prb,amsbsy]{revtex}

\begin{document}
\wideabs{

\title{Small--angle scattering of neutrons on large scale inhomogeneities
in refraction range}

\author{Y. N. Skryabin}
\address{Institute for Metal Physics, Russian Academy of Sciences, \\
Ural Division, 620219,Ekaterinburg, Russia}
\author{A. V. Chukin}
\address{Ural State Technical University,
620002, Ekaterinburg, Russia}

\date{\today}

\maketitle

\begin{abstract}
The small--angle scattering of neutrons on large scale 
inhomogeneities of two--component systems in refraction 
range is considered. We supposed, that neutrons acquire
phase shifts describing by the Gaussian distribution function. 
It is shown that the average cross section 
of scattering contains two contributions. The first
contribution describes the diffraction and in the asymptotic 
limit of large transfer momentum this contribution is 
proportional to $q^{- \Delta}$, 
where $2 \leq \Delta \leq 4$.
The second contribution is similar to the Fraunhofer 
diffraction, which in the asymptotic limit of large transfer 
momentum is proportional to $q^{-3}$.
\end{abstract}
\pacs{}
}

\section{Introduction}

Small--angle neutrons scattering is the powerful method to study large 
scale inhomogeneities of different structures in condensed matter
\cite{weiss,debay,wong,sinha,svergun}. 
Usually, experimental data are interpreted within the Born 
approximation.
However, when neutrons are scattered on large scale inhomogeneities
refraction processes can give a significant contribution 
in the scattering cross section. 
The amplitude of small--angle scattering of neutrons under propagation
through an inhomogeneity is defined by the phase shift
$\delta(\boldsymbol{\rho})$, where 
$\boldsymbol{\rho}$ is a transfer momentum
perpendicular to the direction of the neutron beam (see for example
\cite{landau}). 
In the case, when $\delta(\boldsymbol{\rho}) \ll 1$, the
Born approximation is valid, in the opposite case, when 
$\delta(\boldsymbol{\rho}) \gg 1$,  
refraction effects became considerable and the Born approximation
is not applicable \cite{weiss}.

It is well known \cite{debay} that for two--component system having 
inhomogeneities with smooth boundaries the intensity of scattering 
$I(\mathbf{q})$ falls off with increase of the transfer momentum 
$\mathbf{q}$ according to the Porod law $q^{-4}$.
However in most cases the surface dividing phases of the system 
cannot be considered as smooth. 
In particular, the scattering of neutrons in
Ising--like magnetic systems in an external magnetic field
\cite{wong,mukamel,grinstein,wong83,wong84,birgeneau}
occurs on domain walls, with a surface  strongly differs 
from a smooth plane.
The scattering on two--component systems with
rough interfaces of phases is investigated in detail within
the Born approximation \cite{wong,sinha}. 
It was shown that the intensity of scattering has
the contribution, which is described by the power
function of the transfer momentum with the exponent less than 4. 
Deviations from the Porod law take place also for scattering 
on fractal structures \cite{bale}. 
These results agree well with a lot of experimental data. 
It is necessary to emphasize, that in this case a refraction 
processes are absolutely omitted from consideration.
However these processes can be very important for the studying of 
neutron scattering in systems with large scale 
inhomogeneities when the Born approximation becomes 
inapplicable.

In the theory of the scattering on large scale spherical 
inhomogeneities it is well known \cite{hulst} that for the case 
of an inapplicability of the Born approximation
the amplitude of scattering consists of three terms 
(diffraction, refraction and so--called residual term) 
in correspondence with 
the expression for the forward scattering amplitude   
by spherical inhomogeneities (see for example \cite{landau}). 
However, we don't known the precise expression
for residual term unlike the diffraction and refraction 
contribution to the scattering amplitude, 
but we can to expand this expression in power series
of $1/\alpha$ (where $\alpha$ is the Born parameter) 
and to get approximate estimation only.
According to this fact the range where $\alpha \gg 1$ was 
called refraction, and the range 
$ \alpha \ll 1$ was called diffraction correspondingly.
Recently \cite{skryabin} for the residual term of the amplitude of 
the scattering by spherical inhomogeneities the precise expression 
was obtained. 
This expression contains special functions, the arguments of which 
have two dimensionless parameters: the Born parameter and 
the product $qR$, where $R$ is a radius of the inhomogeneity.
In this case, the transition of the amplitude of scattering from 
the diffraction range to the refraction occurs not for 
$\alpha \sim 1$ but for $\alpha \sim qR$.
It worth to point, that in the asymptotic limit $qR \gg 1$ 
the diffraction range ``analytically continue''
in the area prohibited for it $\alpha > 1$ and, correspondingly, 
the refraction range is shifted.

The influence of the refraction processes in the multiple neutron 
scattering on large scale inhomogeneities is especially 
important \cite{berk,mps}.
According to the general theory of scattering \cite{mott} 
the intensity of the multiple scattering is completely determined 
by single scattering cross section.
Usually, it is considered the case, when the dominating contribution 
to the cross section of single scattering introduces the 
diffraction range of scattering \cite{mt}. 
In this case the cross section of single scattering
on spherical inhomogeneities in the limit $qR \gg 1$ is described 
by the function $(qR)^{-4}$. 
In the refraction range in the same limit $qR \gg 1$
the single cross section of scattering contains terms
$(qR)^{-3}$ and $(qR)^{-4}$ due to the existence of the 
refraction term \cite{weiss}. 
Therefore for the case, when the refraction range
gives the dominant contribution to the cross section of 
single scattering, the change of the intensity of multiple 
scattering take place due to an appearance of new asymptotic 
behavior of single scattering cross section
\cite{mps}.

The main aim of this work is to study just refraction processes
in systems where the declination of the scattering intensity 
from the Porod law is observed. 
Next such systems we call the fractal systems. 
We do not known the amplitude of scattering in the refraction range 
when the scattering occurs in the fractal medium unlike of scattering 
on spherical inhomogeneities.
However the amplitude of scattering in the refraction range
can be restored by the known amplitude of the Born approximation if we 
use the formal expression of phase shift by a Fourier transform of 
the Born amplitude of scattering on transfer momentum. 
It is worth especially to emphasize, that there is no necessity 
to assume the smallness of the Born parameter of the theory of 
scattering in this case. 
Hereinafter we interested by systems with known cross section 
instead of the scattering amplitude in the Born approximation. 
The Born cross section is presented by the correlation
function of local fluctuations of the Born scattering amplitude
\cite{wong} that is local fluctuations of phase shifts. 
The main assumption of our work is that we  average the correlation 
function according to the Gaussian function  
that is we assume that phase shifts are Gaussian random variables 
and mean--square fluctuations of phase shifts coincide with the Born
scattering cross section . 
Thus the correlation functions are easily calculated and the refraction 
part of the cross section can be selected.
The choice a distribution function  by the Gaussian function for the
description of the refraction range gives the right expression for
the well known Babinet principle with exponential accuracy. 
This allows to hope that results obtaining in  
this work are correct in this range of scattering.

\section{Scattering on two--component fractal systems in Born approximation}

Within the limit of small phase shifts 
$\delta(\boldsymbol{\rho}) \ll 1$,
the single scattering is described by the Born approximation.
In this limit the scattering intensity
$I ( \mathbf{q})$ is presented by the correlation function 
of density fluctuations of scattering objects
$\gamma(\boldsymbol{\rho})$ \cite{weiss,debay,wong,sinha}.
If for two--component systems the surface is smooth then 
the Porod law for the intensity \cite{debay} occurs, $I(q) \sim q^{-4}$, 
in the limit of large transfer momenta $q$.
If the surface is rough it was obtained by two
various methods \cite{wong,sinha} that the correction to the Porod law 
can be written as  
\begin{equation}
\frac{d\sigma}{d\Omega} \sim \frac{A_1}{q^4}+\frac{A_2}{q^{x+3}},
\label{rough}
\end{equation}
where $x$, $(0 \leq x \leq 1)$ is the parameter describing 
roughness of surface \cite{grinstein}. 
It is necessary to note that in the case
when the second term in Eq. (\ref{rough}) dominates, the
Eq. (\ref{rough})  describes the scattering cross section  
in fractal medium \cite{bale}.
We remind that for scattering on volume fractals the 
scattering cross section within the limit of large transfer 
momenta is proportional to $q^{-D_v}$, where $D_v < 3$, 
for scattering on surface fractals the cross section is
proportional to $q^{-(6-D_s)}$, where $D_s < 3$ and at last 
for a scattering on critical fluctuations the cross section 
is proportional to $q^{-2}$.
Thus the general expression of the scattering cross section 
covering all these particular cases can be described by the equation 
$q^{-\Delta}$, where $2 \leq \Delta \leq 4$.

\section{The cross section in eikonal approximation}

Now we attempt to go beyond the Born case and to study refraction 
processes of small--angle scattering of neutrons in fractal 
systems  using the eikonal approximation \cite{landau,mott}. 
In this case it supposed, that the energy of incident particles is 
a lot of more then the potential energy of scattering and 
the amplitude is written as
\begin{equation}
f({\mathbf{q}})=-\frac{{\mathrm{i}} k} {2\pi} \int (\exp [2{\mathrm{i}}
\delta(\boldsymbol{\rho})]-1) \exp (-{\mathrm{i}}{\mathbf{q}}
\boldsymbol{\rho}) d^2\boldsymbol{\rho},
\label{amplitude}
\end{equation}
where $\mathbf{q}$ is the two--dimensional transfer momentum lying in the
plane perpendicular to the beam of neutrons with momentum $k$,
$\delta(\boldsymbol{\rho})$ is the phase shift.
The cross section is defined as
\begin{eqnarray}
\frac{d\sigma}{d\Omega} & = & |f({\mathbf{q}})|^2 
= \Bigl(\frac{k}{2\pi}\Bigr)^2 \int d^2 \boldsymbol{\rho}
d^2\boldsymbol{\rho}'
\exp(-{\mathrm{i}}{\mathbf{q}}(\boldsymbol{\rho} -
\boldsymbol{\rho}')) \times \nonumber\\
& &\Bigl( S(\boldsymbol{\rho}) - 1 \Bigr)
\Bigl( S^{\ast}(\boldsymbol{\rho}') - 1 \Bigr) \nonumber\\
&=&\Bigl(\frac{k}{2\pi}\Bigr)^2 \int d^2 \boldsymbol{\rho}
\exp(-{\mathrm{i}}{\mathbf{q}}
\boldsymbol{\rho}) \gamma '(\boldsymbol{\rho}),
\label{cross}
\end{eqnarray}
here we define the function
\begin{equation}
\gamma '(\boldsymbol{\rho}) = \int d^2\boldsymbol{\rho}'
\Bigl( S(\boldsymbol{\rho}' + \boldsymbol{\rho}) - 1 \Bigr)
\Bigl( S^{\ast}(\boldsymbol{\rho}') - 1 \Bigr),
\label{correl}
\end{equation}
and also
\begin{equation}
S(\boldsymbol{\rho})=\exp 2{\mathrm{i}}\delta(\boldsymbol{\rho}),
\quad
\delta(\boldsymbol{\rho})=-\frac{1}{2\hbar v}
\int U(\boldsymbol{\rho},z)dz,
\label{smatrix}
\end{equation}
$U(\boldsymbol{\rho},z)$ is the scattering potential, and $v$ is the 
velocity of particles. For small phase shifts the exponential
functions in Eqs. (\ref{correl}, \ref{smatrix}) can be expand in 
power series and the correlation function $\gamma '(\boldsymbol{\rho})$
is transformed to the $\gamma (\boldsymbol{\rho})$ of the Born approximation.

It is easy to see from Eq. (\ref{cross}) that the scattering cross section is
determined by the dependence of phase shift $\delta$ on the impact 
parameter $\boldsymbol{\rho}$.
This dependence can be find by the known Born scattering amplitude. 
It should be noted that formally the phase shift is the Fourier transform
of the scattering amplitude in the Born approximation
\begin{equation}
\delta (\boldsymbol{\rho}) = \frac{\pi}{k} \int \frac{d^2 {\mathbf{q}}}
{(2 \pi)^2} f_B({\mathbf{q}}) \exp ({\mathrm{i}}{\mathbf{q}}
\boldsymbol{\rho}).
\label{shift}
\end{equation}
In other words instead of using the direct definition of the phase shift
by the scattering potential Eq. (\ref{smatrix})
it can be more suitable sometimes (for example at the scattering on spherical
inhomogeneities or superconducting vortex lines) to find the phase
shift by the known Born amplitude.
Besides this method allows in the eikonal approximation analytically to
continue the phase shift to the region of large Born parameters since
the formal equation (\ref{shift}) does not assume a smallness of this
parameter in the scattering theory.

Unfortunately we known the Born cross section but don't known 
the Born amplitude for fractal medium. 
However the cross section is represented by the
appropriate density fluctuations correlation function 
(see for example \cite{landau2}). 
To present the cross section in refraction range by the
correlation function of density fluctuations 
we assume that phase shift is a local random variable 
which in the primary approximation can be
described by the Gaussian distribution. 
Define the Gaussian distribution function
of phase shifts  $\delta_{\mathbf{q}}$ in
the momentum representation $\delta_{mathbf{q}}$ as 
\begin{equation}
P(\delta_{\mathbf{q}}) = \frac1Z \exp (-\frac {|\delta_{\mathbf{q}}|^2}
{2 \langle |\delta_{\mathbf{q}}|^2 \rangle}),
\label{gauss}
\end{equation}
where $Z$ the is the partition function written by the functional integral
\begin{equation}
Z = \int {\mathcal{D}} \delta_{\mathbf{q}}
\exp (-\frac {|\delta_{\mathbf{q}}|^2}
{2 \langle |\delta_{\mathbf{q}}|^2 \rangle}).
\label{partition}
\end{equation}
The mean--square fluctuation of phase shifts defines the Born cross
section of scattering in the form
\begin{equation}
\langle |\delta_{\mathbf{q}}|^2 \rangle = \frac1S \Bigl( \frac{\pi}{k}
\Bigr)^2 \langle \frac{d\sigma}{d\Omega} \rangle_B.
\label{meansq}
\end{equation}
Here $S$ is the geometrical cross section of inhomogeneity and
$\langle \frac{d\sigma}{d\Omega} \rangle_B $ is the Born cross
section of single scattering. The angle brackets means the average over
the Gaussian fluctuations. 

For the cross section we have
\begin{eqnarray}
\langle \frac{d\sigma}{d\Omega} \rangle = \Bigl( \frac{k}{2\pi} \Bigr)^2
\int d^2\boldsymbol{\rho} d^2\boldsymbol{\rho}'
\exp (-{\mathrm{i}}{\mathbf{q}}(\boldsymbol{\rho} -
\boldsymbol{\rho}')) \times \nonumber\\
\langle (\exp (2{\mathrm{i}} \delta(\boldsymbol{\rho})) - 1)
(\exp (-2{\mathrm{i}} \delta(\boldsymbol{\rho}')) -1) \rangle.
\label{dif}
\end{eqnarray}
Calculating the mean value of exponential functions of random variables
distributed according to the Gaussian function \cite{landau2}, we
rewrite expression Eq. (\ref{dif}) as
\begin{eqnarray}
\langle \frac{d\sigma}{d\Omega} \rangle & = &\Bigl(\frac{k}{2\pi}\Bigr)^2
\int d^2 \boldsymbol{\rho} d^2\boldsymbol{\rho}'
\exp(-{\mathrm{i}}{\mathbf{q}}(\boldsymbol{\rho}-
\boldsymbol{\rho}')) \times \nonumber\\
& &\Biggl\{ \exp \Bigl[ - \frac4S \Bigl( \frac{\pi}{k} \Bigr)^2 \int
\frac{d^2 {\mathbf{q}'}}{(2\pi)^2}
\langle \frac{d\sigma}{d\Omega} \rangle_B \times \nonumber\\
& &\bigl( 1 - \exp({\mathrm{i}}{\mathbf{q}'}
(\boldsymbol{\rho - \rho}')) \bigr) \Bigr] \nonumber\\
& & - 2 \exp \Bigl( - \frac2S \Bigl( \frac{\pi}{k} \Bigr)^2 \int
\frac{d^2 {\bf q'}}{(2\pi)^2}
\langle \frac{d\sigma}{d\Omega} \rangle_B \Bigr) + 1 \Biggr\}.
\label{dif2}
\end{eqnarray}
Expanding of exponents in power series in Eq. (\ref{dif2}) leads
to the Born cross section that corresponds to transition in the
diffraction range of scattering.

It is conveniently to define the function $\sigma_B(\rho)$ as
\begin{eqnarray}
\sigma_B(\boldsymbol{\rho}) & = & \frac{1}{k^2} \int d^2{\mathbf{q}}
\langle \frac{d\sigma}{d\Omega} \rangle \exp ({\mathrm{i}}{\mathbf{q}}
\boldsymbol{\rho}), \nonumber\\
\sigma_B(\rho = 0) & = & \sigma_B(0).
\label{sigma}
\end{eqnarray}
Here $ \sigma_B(0) $ is the total Born cross section where the magnitude
of the Born parameter is large in refraction range of scattering.
Substituting Eq. (\ref{sigma}) in Eq. (\ref{dif2}) we have
\begin{eqnarray}
\langle \frac{d\sigma}{d\Omega} \rangle & = & \frac{k^2}{2\pi}S
\int d \rho \rho J_0(q\rho)
\exp \Bigl( - \frac{\sigma_B(0)}{S} \zeta(\rho) \Bigr) \nonumber\\
& & + \Biggl( 1 - 2\exp (-\frac{\sigma_B(0)}{2S}) \Biggr)
\Bigl( \frac{k\xi}{q} \Bigr)^2J_1^2(q\xi),
\label{dif3}
\end{eqnarray}
where we define the function
\begin{equation}
\zeta(\rho) = 1 - \frac{\sigma_B(\rho)}{\sigma_B(0)}.
\label{zeta}
\end{equation}
The first term in Eq. (\ref{dif3}) is the known expression for the scattering
intensity in the theory of multiple neutron scattering \cite{mt}.
It is necessary to emphasize, that in this case the value of
$\sigma_B(0)/S$ in refraction range became large in contrast with unit
more just due to that in the equation for $\sigma_B(0)$ the smallness of
the Born parameter does not assume. 
Calculating the integral in the last term in Eq. (\ref{dif3}) we assume
that a size of fractal objects under scattering is limited by correlation
length $\xi$ due to average by all chaotic orientations
of the fractal object of course. It worth to note that the obtained
expression for the second term is analogous to the Fraunhofer diffraction
on the round hole \cite{landau3}.
In the asymptotic limit of large transfer momenta the Fraunhofer
diffraction is described by the function  $q^{-3}$.
For an arbitrary fractal the result of the calculation of this integral
would be similar of the Fraunhofer diffraction on the hole which has a form
corresponding to this fractal.

Calculating the integral Eq. (\ref{dif2}) over the two--dimensional vector 
$q$, we have the total cross section
\begin{equation}
\sigma = \int d{\mathbf{q}} \langle \frac{d\sigma}{d\Omega} \rangle =
2S \Bigl[1 - \exp \Bigl(-\frac{\sigma_B(0)}{2S} \Bigr) \Bigr]
\label{total}
\end{equation}
The ratio $\sigma_B(0)/2S$ is proportional to the square of the Born parameter.
The Eq. (\ref{total}) can be rewrite as the equation in the Born
approximation if the Born parameter is small in comparison with unit. 
The total cross section is equal to the double geometrical section in 
the case of the large Born parameter. 
This result is in complete correspondence with the Babinet principle. 
According to this principle the total cross section is
determined by a double cross section of an absorption \cite{landau,landau3}.

In general the differential cross section of scattering in the Born 
approximation can be presented as
\begin{equation}
\langle \frac{d\sigma}{d\Omega} \rangle_B = Aq^{-\Delta}f_{\Delta}(q\xi),
\label{born}
\end{equation}
where $A$ is a constant. $\Delta$ is a some parameter in the interval
$(2 \leq \Delta \leq 4)$. The function $f_{\Delta}(q\xi)$  within the
limit of a large $q\xi$ can be replace by unit but for the case of small
transfer momentum it ensures not singular behavior for the scattering cross 
section. Substituting Eq. (\ref{born}) in Eq. (\ref{zeta}) and making the
substitution of a variable $\lambda=k\rho$, we get
\begin{equation}
\zeta(\lambda) = \frac{\int_{0}^{\infty}dxx^{- \Delta +1}
[1 - J_0(\vartheta_0\lambda x]
f_{\Delta}(x)}{\int_{0}^{\infty}dxx^{-\Delta + 1}f_{\Delta}(x),
\label{zeta2}}
\end{equation}
where  $\vartheta_0 = 1/k \xi$.

Since the main contribution to the integral of Eq. (\ref{dif3}) is given by
small  $\lambda$, we can replace $f_{\Delta}(x)$  by unit in
the numerator of Eq. (\ref{zeta2}). As a result for  $2<\Delta<4$
we have a known integral, which results to
\begin{equation}
\zeta(\lambda) = (\vartheta_0\lambda/2)^{\Delta - 2}/F_{\Delta},
\label{zeta3}
\end{equation}
where
\begin{equation}
F_{\Delta} = \frac{(\Delta - 2) \Gamma(\Delta /2)}{\Gamma (2 - \Delta /2)}
\int_{0}^{\infty}dxx^{1 - \Delta}f_{\Delta}(x).
\label{fdelta}
\end{equation}
The equation  (\ref{dif3}) is:
\begin{eqnarray}
\langle \frac{d\sigma}{d\Omega} \rangle & = &\frac{S}{2\pi}\frac{k^2}{q^2}
\int_{0}^{\infty}dxx
J_0(x) \exp \Bigl[ - \Bigl( \frac{\sigma_B}{SF_{\Delta}} \Bigr) \Bigl(
\frac{x}{2q\xi} \Bigr)^{\Delta - 2} \Bigr] \nonumber\\
& & +\Bigl[ 1 - 2 \exp \Bigl( - \frac{\sigma_B}{2S} \Bigr) \Bigr] \Bigl(
\frac{k\xi}{q} \Bigr)^2J_1^2(q\xi).
\label{x}
\end{eqnarray}
Here we transformed the first term in Eq. (\ref{x}) similar to the known
equation \cite{maleyev}, therefore it is possible to use known result
\cite{mott} that in asymptotic limit the intensity of multiple scattering
coincide with the intensity of single scattering. In this case the
single scattering intensity is defined by Eq. (\ref{born}).
The refraction term  (which in asymptotic limit is proportional to $q^{-3}$),
can become the main one in the intensity depending on a kind of the law
for the intensity of a single scattering. Really, for $3 < \Delta < 4$ exists
a range of values of transfer momentum where in the intensity
dominates the Fraunhofer diffraction with asymptotic dependence
$q^{-3}$. The range $3 < \Delta < 4$
corresponds to the scattering on surface fractals when the parameter  $x$
in the Eq. (\ref{rough}) change in limits $0 < x < 1$. In the other limits
$2 < \Delta < 3$, which corresponds to the scattering on a
volume fractals, the Fraunhofer diffraction already does not render
essential influence to the intensity of scattering already.

In the specific case $\Delta = 4$ the cross section Eq. (\ref{born}) is 
described by the Porod law. 
This case is well investigated in the theory of multiple
scattering on spherical inhomogeneities \cite{mt,berk}. Here the
intensity of multiple scattering transforms in the intensity of single
scattering in asymptotic limit of large transfer momentum too \cite{mott}.
The analysis of this case begins with the equation (\ref{dif3}),
where for the first term we can use the result of the
multiple scattering theory again. This result shows, that in an asymptotic
limit the intensity of multiple scattering is determined by the cross
section in the Born approximation $q^{-4}$. Therefore, for this  case, as
well as for previous, it is possible to assert, that the cross section of
small-angle neutron scattering in refraction range in fractal medium
with a parameter ($x=1$ in the equation (\ref{rough})) is described
by the Fraunhofer diffraction.

The refraction term does not render essential influence to the intensity
of scattering in the other special case, when  $\Delta=2$. This case
describes the  scattering on critical fluctuations. Here also it
is necessary to analyze the Eq. (\ref{dif3}) and we must for the first
term to use the result of the multiple scattering theory. 
This  multiple scattering theory on
critical fluctuations is considered in Ref.\cite{mt}, where the dependence
of the multiple scattering intensity on the transfer momentum in the
asymptotic limit of large momenta was found. 
These functions give the predominant
contribution in comparison with the refraction term.

\section{Conclusions}

Thus at studying of the small--angle neutron scattering in fractal medium 
in the refraction range (that is for the case when the Born parameter of the
theory of scattering is not small) it is necessary for the single cross
section to take into account the contribution proportional $q^{-3}$  
within the large transfer momentum limit which is similar to the  Fraunhofer
diffraction. 
Depending on the value of the exponent $\Delta$  
of the single scattering cross section in the Born approximation 
the averaged correlation function over the Gaussian random variables in 
refraction range gives a very small contribution to the cross section for 
$\Delta > 3$. 
In the opposite case the Fraunhofer diffraction is negligible.
\acknowledgments

This work was partially supported by Russian Foundation for Basic Research
under grant number 97-02-17315.

\end{document}